\documentclass[conference]{IEEEtran}
\IEEEoverridecommandlockouts
\usepackage{cite}
\usepackage{amsmath,amssymb,amsfonts}
\usepackage{algorithmic}
\usepackage{graphicx}
\usepackage{textcomp}
\usepackage{xcolor}
\usepackage{booktabs} % For formal tables
\usepackage{ifthen}
\usepackage{graphicx}
\usepackage{subfigure}
\usepackage{longtable}
\usepackage{listings}
\usepackage{enumitem}
\usepackage{indentfirst}
\usepackage{diagbox} 
\usepackage{mwe}
\usepackage{url}
\usepackage{hyperref}
\usepackage{multirow}

\def\BibTeX{{\rm B\kern-.05em{\sc i\kern-.025em b}\kern-.08em
    T\kern-.1667em\lower.7ex\hbox{E}\kern-.125emX}}
\begin{document}

\title{Exploring Trade-offs in Dynamic Task Triggering for Loosely Coupled Scientific Workflows\\
}

\author{\IEEEauthorblockN{Zhe Wang,
Pradeep Subedi,
Shaohua Duan,
Yubo Qin,
Philip Davis,
Anthony Simonet,
Ivan Rodero,
Manish Parashar}
\IEEEauthorblockA{\textit{Rutgers Discovery Informatics Institute, Rutgers University,} \\
Piscataway, USA \\
\{jay.wang,pradeep.subedi,parashar\}@rutgers.edu}

}

\maketitle

\begin{abstract}
In order to achieve near-time insights, scientific workflows tend to be organized in a flexible and dynamic way. Data-driven triggering of tasks has been explored as a way to support workflows that evolve based on the data. However, the overhead introduced by such dynamic triggering of tasks is an under-studied topic. This paper discusses different facets of dynamic task triggers. Particularly, we explore different ways of constructing a data-driven dynamic workflow and then evaluate the overheads introduced by such design decisions.
We evaluate workflows with varying data size, percentage of interesting data, temporal data distribution, and number of tasks triggered. Finally, we provide advice based upon analysis of the evaluation results for users looking to construct data-driven scientific workflows.
\end{abstract}


\section{Introduction}
\label{sec:intro}
The basic components of a scientific workflow include data producers, such as a physical simulation; data consumers, such as analytics or visualizations; and middleware, such as an I/O service or workflow management tools. One of the fundamental research challenges of scientific workflows is to solve the imbalance between the available I/O bandwidth and the computational capability of underlying hardware~\cite{childs2012situ}. The mismatch between the growth rate of I/O bandwidth and computation frequently results in I/O being the bottleneck of scientific workflows~\cite{kress2016preparing}. Thus, various research efforts have aimed to employ in-situ/in-transit processing, where the data generated by simulation is either consumed in place or transferred to a storage device in near-real-time~\cite{pugmire2016visualization,kress2016preparing,kress2017situ,marrinan2018transferring}.

In-situ/In-transit data processing techniques are designed to improve resource uitilization during data transfer or consumption in order to improve the performance of the workflow. These workflows are also being organized in a dynamic fashion to increase the flexibility. The main idea of dynamic workflows for scientific applications is to make decisions about operations such as task start/stop/modification according to the content of the data or events in the workflow rather than using a predefined execution sequence~\cite{salloum2015enabling,mattoso2015dynamic,bennett2016trigger}. While scientific workflows can be classified as either tightly-coupled or loosely-coupled based upon the task composition and deployment \cite{kress2019comparing}, in this paper we focus on loosely coupled workflows. %which represent tasks by separate programs, as there are natural advantages to implement a dynamic task trigger in a decoupled workflow composition.
% todo add more trigger citations

When designing a dynamic workflow, one of the foremost challenges is how to construct a workflow that evolves based on the data content. The dynamic task trigger mechanism is one of the most commonly used approaches~\cite{bennett2015sublinear,ling2017using,insitutriggure} to solve this problem. At a high level, the services required for supporting the dynamic task trigger mechanism can be divided into two parts: a data checking service that decides when to run a task, and an associated task operator that determines how to run the task~\cite{bennett2016trigger,insitutriggure}. The main research challenge of designing the data checking service is how to use domain-specific metrics to identify data characteristics~\cite{bennett2016trigger,ling2017using}. Meanwhile, the main research challenge of designing the trigger-based task operator is how to improve the trigger flexibility with minimal overhead based on the capabilities of underlying middleware~\cite{jin2012scalable,perarnau2015distributed,pandey2018event,insitutriggure}.

Although dynamic task triggering can improve the flexibility of workflow management, it also introduces overhead. For example, the process of checking data content or scheduling new tasks during the workflow may cause an extra delay and slow down workflow execution. Different service designs for task triggering carry different trade-offs. For example, the data checking service can be bound with the data producer, the data consumer, or the data staging service included by the middleware. To the best of the authors’ knowledge, few works have explored this overhead and analyzed the underlying factors that influence the overhead. Without understanding reasons that cause the overhead, it is difficult to optimize the dynamic task trigger service and adapt it to workflows with various initial settings. 

The goal of this paper is to analyze the trade-offs between the benefits and the overhead of dynamic task triggers of the loosely coupled workflow. Specifically, this paper aims to illustrate the factors that influence the performance of a workflow that contains dynamic task triggers. By evaluating these factors, this paper also aims to explore the underlying reasons that cause these trade-offs. These reasons can lead to best practices that maximize the benefits and minimize the overhead of dynamic task triggering for scientific workflows. The main contributions of this paper are:

\begin{itemize}[noitemsep,leftmargin=*]
\item We address typical dynamic task trigger patterns for loosely-coupled scientific workflows, systematizing according to the design choice of the task checking service
\item We evaluate how various workflows' initial settings impact the performance of the workflows containing dynamic task triggers.
\item We analyze the factors that influence the overhead of dynamic task triggers based upon experimental results and present research opportunities to maximize the benefits and minimize the overhead of a scientific workflow using dynamic task triggers.
\end{itemize}

The remainder of the paper is organized as follows. Section~\ref{sec:relatedwk} discusses related work. Section~\ref{sec:anaysis} presents different patterns for implementing the dynamic task triggers. Section~\ref{sec:experiment} illustrates the details and results of the experiments. Section~\ref{sec:conclusion} concludes this paper and discusses remaining research challenges.

\section{Related Work}
\label{sec:relatedwk}

Recent research has classified trigger mechanisms according to the semantics of triggers, namely the domain-specific and the domain-agnostic~\cite{insitutriggure}. Besides, the trigger mechanisms can also be summarized from the perspective of workflow construction. In particular, Directed Acyclic Graph (DAG) and non-DAG~\cite{yu2005taxonomy} are two typical types of the workflow structures. We use the term \emph{static trigger} to represent the mechanism of constructing workflow based on the predefined DAG configuration. Similarly, \emph{dynamic trigger} represents a workflow that operates tasks based on the data content using a non-DAG pattern.

\subsection{Static Task Triggers}

A static trigger system utilizes a predefined task execution sequence to construct the task dependencies before workflow execution. This sequence is usually represented by a DAG between tasks. Workflow management tools that provide static trigger support~\cite{albrecht2012makeflow,wilde2011swift,deelman2015pegasus,ludascher2006scientific} construct the dependencies between tasks using configuration files or a domain-specific language. The conditions on which to trigger the subsequent operations are the completion of dependent tasks. However,  this static approach lacks the ability to control the status of tasks based upon the content of generated data during workflow execution. For example, the cosmological simulation HACC~\cite{habib2016hacc} generates petabyte-scale data during execution, but only the data that contains specific physical phenomenon, such as a halo shape, are meaningful to scientists. In addition, it is difficult to know in advance when this phenomenon will appear, so static task triggers fail to express the task trigger condition for this type of workflow. Based on a survey of workflow management tools~\cite{da2017characterization}, managing scientific workflows using a data-driven pattern is one of the challenges that needs to be solved by future workflow tools, especially for composing the extreme-scale applications.

% update expression
%It is necessary to explore how the flexible workflow composition support trigger mechanisms based on the content of the data, especially for extreme-scale applications~\cite{da2017characterization}.

\subsection{Dynamic Task Triggers}

Dynamic task triggers control the status of the tasks based on the content of data during workflow execution rather than using a fixed pre-defined task execution sequence. Research works have proposed various approaches to designing dynamic task triggers. Larsen et al.~\cite{insitutriggure} integrated user-defined triggers into a visualization service. The decision of when to trigger tasks may come from the mesh topology, scalar fields, or performance data. Jin et al.~\cite{jin2012scalable} used a publish/subscribe (pub/sub) mechanism to implement a messaging middleware and integrate it within an in-memory storage service to trigger tasks. In addition, Pandey et al.~\cite{pandey2018event} designed an ensemble manager to manage the user-defined triggers via monitoring the modification of a specific file or directory; however, these works mainly focus on how to initiate tasks based on various dynamic trigger mechanisms, while few of them discuss the overhead of the dynamic triggers and how workflow configuration influences this overhead.

% todo reorganize this part, delete the description of the 3.1 and combine it with the 4.1 consider to put the introduction at the beginning ??? background ??? put the motivation example at here ??? motivation application, different types of the organization, the analysis of the cases, The contents need to be simplified for the first paragraph at the experiment part, controllable, domain scientics, use miniapp to show this. Analyse the same underlying structure, so why there are those components and the checking service is important, which is in vis, which is in middleware, which is in sim. (the weak point in sim is that there is not global view for data) still focuse on the data checking, if data checking is in the middleware, there is no obvious difference to use the in-memory or by disk, list three examples, at the simulation, thethe bernnet's work, the ascent, at the consumer, the catalyst, visual toool at the middleware, the duan's work, data stager's work

\section{Dynamic Task Trigger Patterns}
\label{sec:anaysis}

In this section, we first introduce a motivating application workflow and explain how the dynamic task trigger mechanism optimizes this workflow. Then, we summarize three typical dynamic trigger patterns based on different design strategies of checking the content of simulation data.

\subsection{Motivating Application Workflow}
\label{sec:moti}

\begin{figure}[t]
\centering     %%% not \center
\subfigure[]{\label{fig:gsa}\includegraphics[width=30mm]{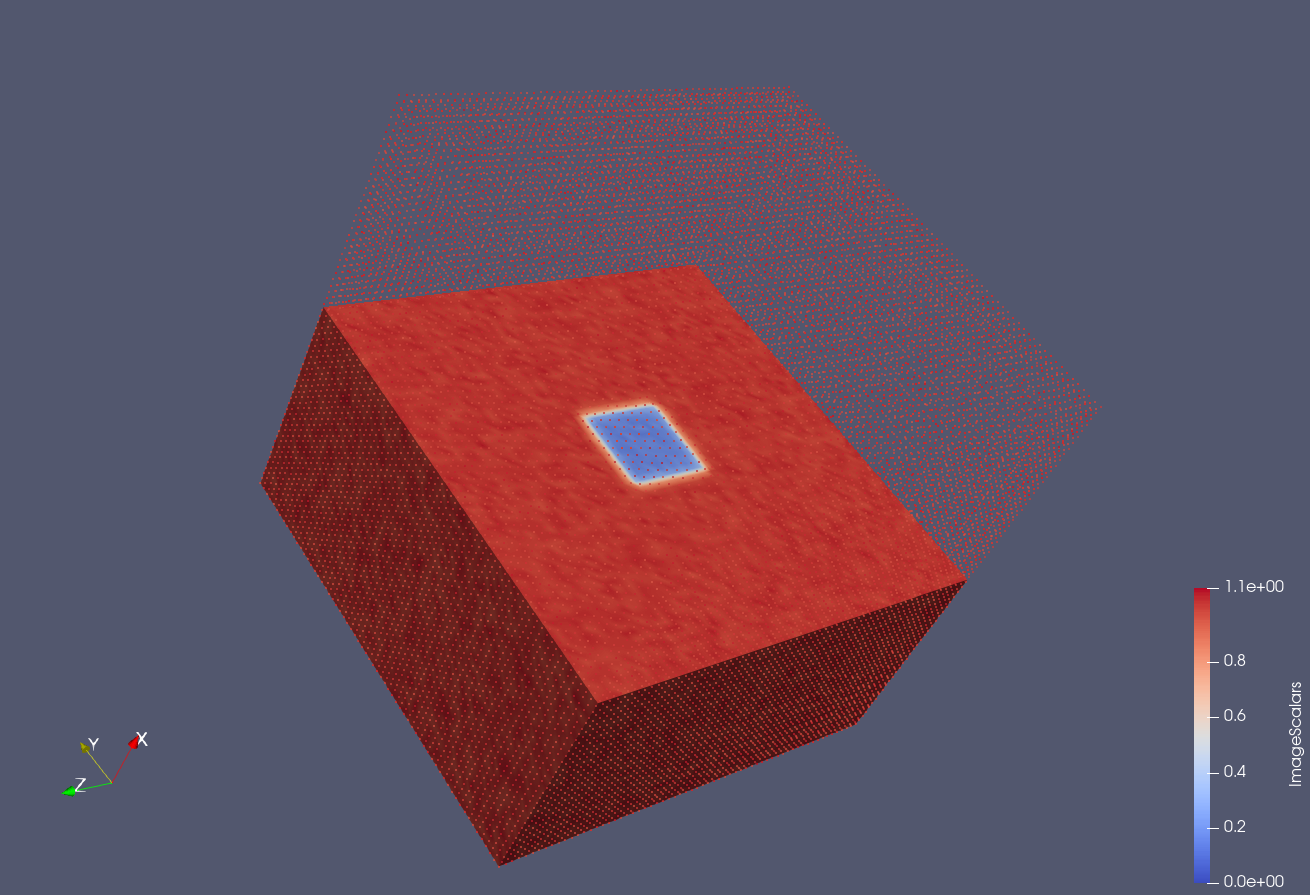}}
\hspace{0.3in}
\subfigure[]{\label{fig:gsb}\includegraphics[width=30mm]{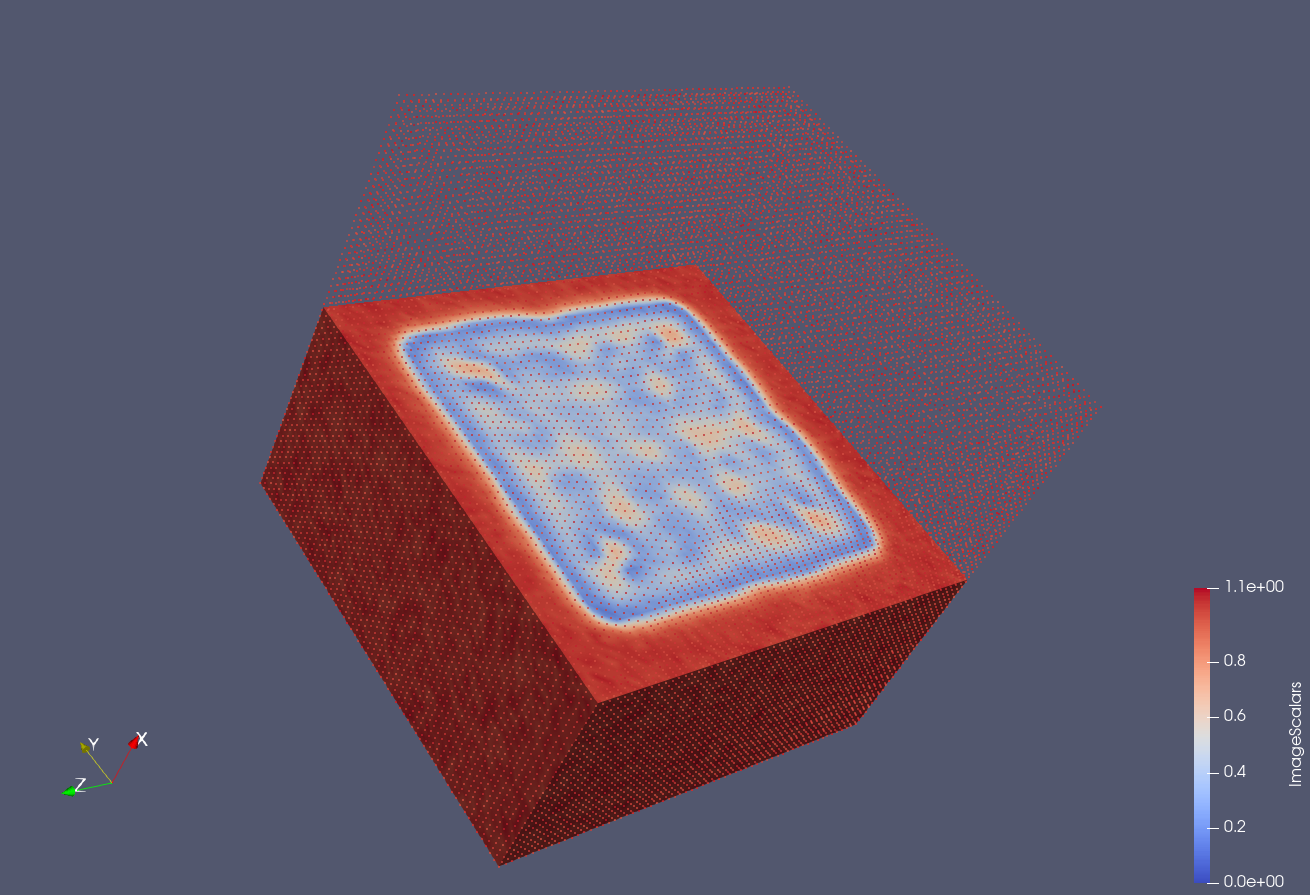}}
\caption{The sub-figure (a) is the visualization with concentrated blue color region. The sub-figure (b) is the visualization with scattered blue color region.}
\label{fg:grayscottsim}
\end{figure}

\begin{figure}[t]
\centering     %%% not \center
\subfigure[]{\label{fig:raw25}\includegraphics[width=20mm]{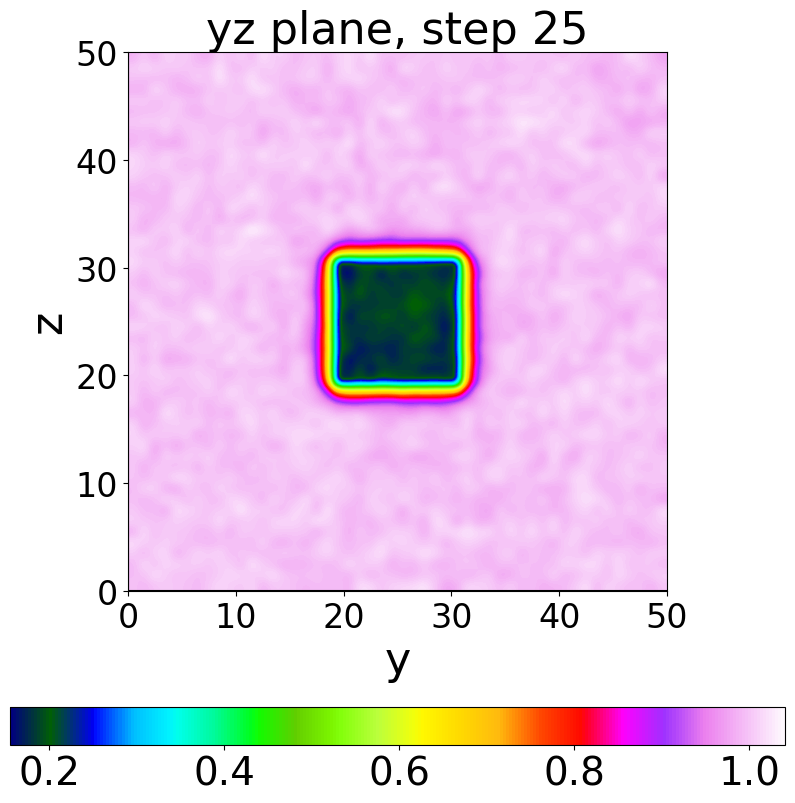}}
\subfigure[]{\label{fig:raw1000}\includegraphics[width=20mm]{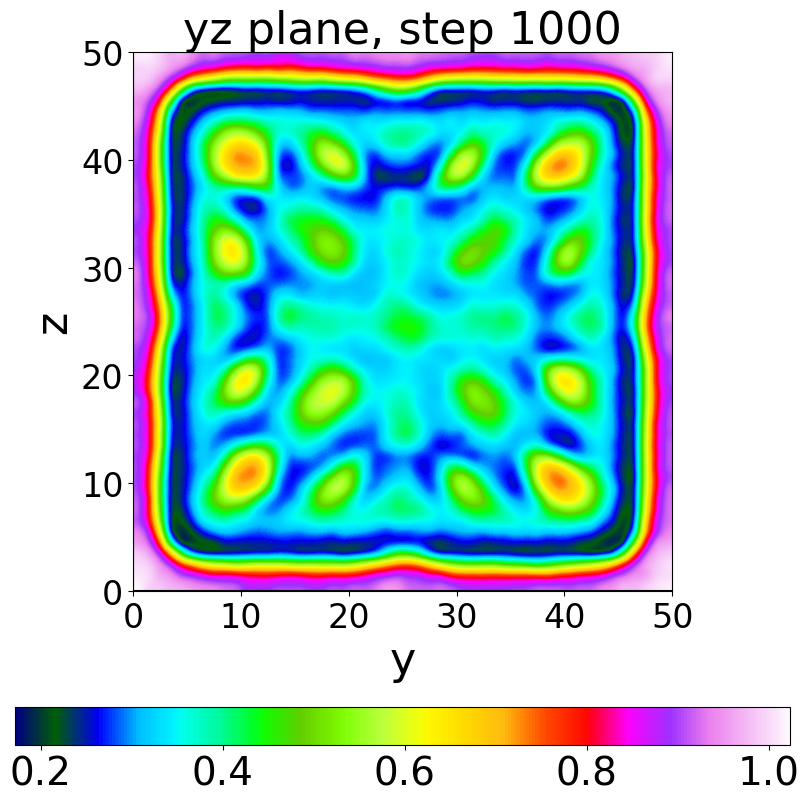}}
\subfigure[]{\label{fig:pdf25}\includegraphics[width=20mm]{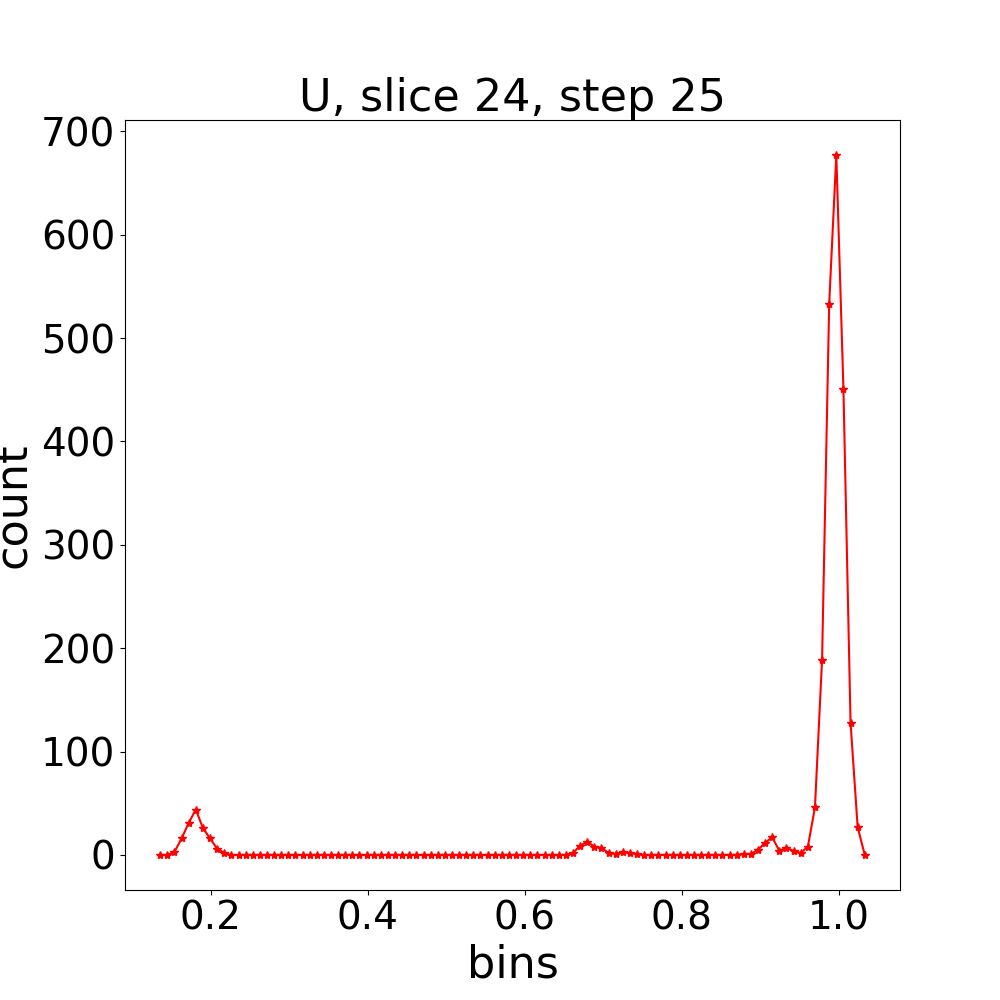}}
\subfigure[]{\label{fig:pdf1000}\includegraphics[width=20mm]{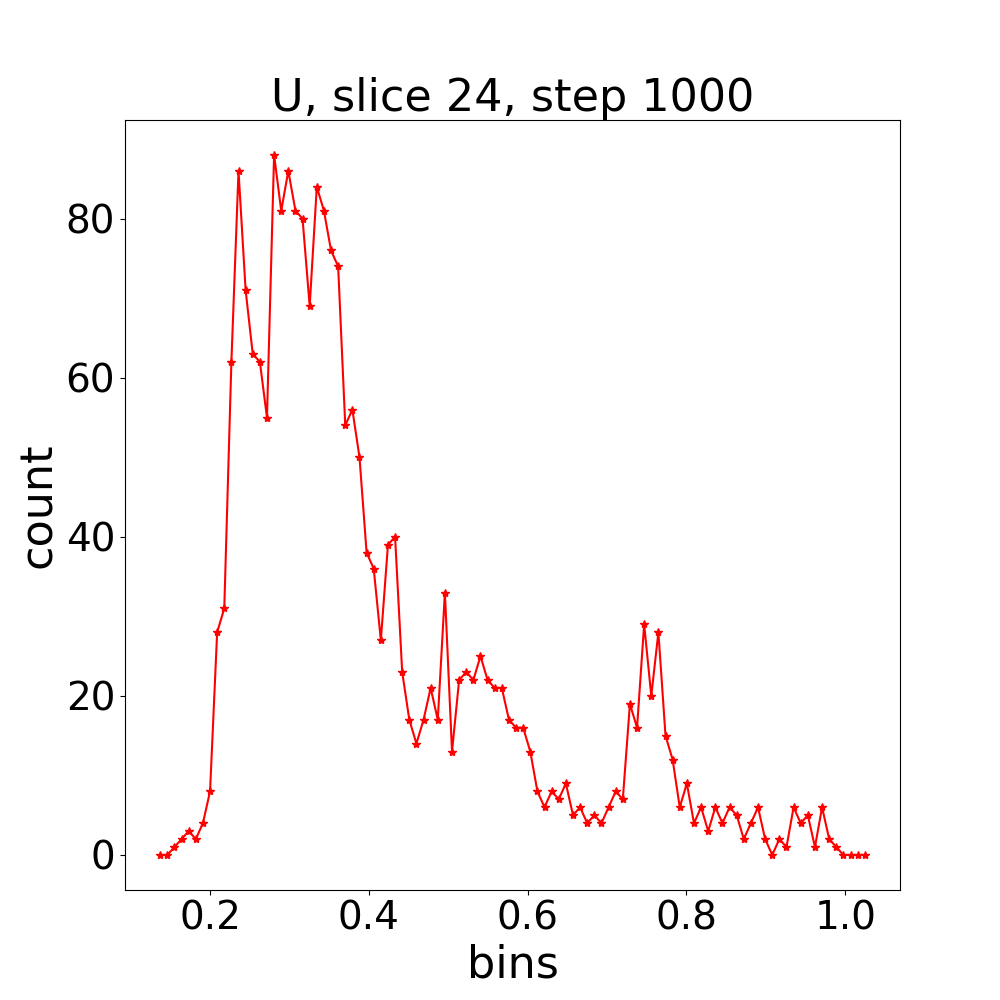}}
\caption{2D planes of the Gray-Scott simulation data and corresponding histograms. The sub-figure (a) is the \textit{yz} plane at step 25 of the simulated data. The sub-figure (b) is the \textit{yz} plane at step 1000 of the simulated data. The sub-figure (c) is the histogram of the simulated variable at step 25. The sub-figure (d) is the histogram of the simulated variable at step 1000.}
\label{fg:clippdf}
\end{figure}

Gray-Scott simulations~\cite{grayscott} reveal a variety of spatiotemporal patterns based on a reaction-diffusion model. One typical analysis for Gray-Scott simulation is to distinguish the pattern of the simulation data. For example in the Figure~\ref{fg:grayscottsim}(a), the blue color region is concentrated at the center region of the visualization; however, in the Figure~\ref{fg:grayscottsim}(b), the blue color region is scattered among whole simulated domain. One commonly used tool to analyse the simulation data is the histogram~\cite{dos2017histogram}. The Figure~\ref{fg:clippdf} shows the colored 2D plane and the corresponding histogram of the simulated variable. It is worth to note the relationship between the curve of the histogram and the scattering level of the simulated data. For example, there is a peak of the curve at the 1.0 in Figure~\ref{fg:clippdf}(c) when the simulated data is concentrated at the center region (Figure~\ref{fg:clippdf}(a)); however, the peak of the curve is located at 0.3 in Figure~\ref{fg:clippdf}(d) when the data is scattered among all space, such as plane shown in Figure~\ref{fg:clippdf}(c). Therefore, the analysis such as calculating peak positions of the histogram can be used to distinguish the pattern of simulated data with lightweight computation, and these metrics are also termed by the \textit{indicator}~\cite{bennett2016trigger}.

% in some complecated simulation such as the weather model, the appearance and the shape of the cloud, same patterns
The process of calculating the \textit{indicator} is an crucial stage to construct the data-driven workflow. In particular, lightweight analytics (\textit{indicator}) are calculated at each step after the generation of simulation data, and time-consuming analytics or other costly operations, such as data dump, are only executed when the indicator value satisfies the predefined constraints during workflow execution. There are several strategies to manage the data checking service (the service that calculates the indicator value) in a scientific workflow. The data checking service can be the in-situ data analytics, which are linked together with the simulation, and the raw data were inspected in memory. This pattern is adopted by the work such as Ascent~\cite{larsen2017alpine} and in-situ trigger detection for S3D simulation~\cite{bennett2016trigger}. Another strategy is to run the data checking service with the data consumer. For example, the Catalyst~\cite{ayachit2015paraview} exports the simulation data to the visualization service. The interesting data can be detected at the Paraview by various filters, and only the interesting data region is visualized. Besides, the data checking service can also run together with the dedicated middleware, such as the data management service. The analytics such as error detection can be integrated with the data staging service~\cite{duan2019addressing} to control the workflow. For example, once a silent error is detected at the staging service before check-pointing, the simulation is rolled back to the last checkpoint and re-executed. 

We use the term~\textit{dynamic task trigger} to represent the process of the data checking and the task triggering during the workflow execution. Based on current solutions discussed above about integrating~\textit{dynamic task trigger} in a scientific workflow, we classify the dynamic task trigger pattern based on the position of data checking services. In particular, the data checking service can be executed at a data producer, consumer, or separate middleware. 

\subsection{Typical dynamic task trigger patterns}

\label{sec:ttp}
\begin{figure*}
\centering     %%% not \center
\subfigure[Producer-responsible]{\label{fg:tcp}\includegraphics[width=55mm]{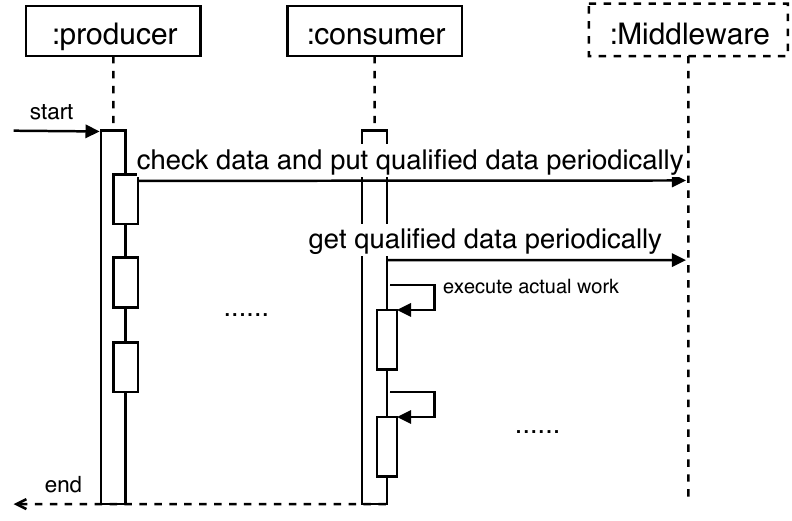}}
\vspace{0.1in}
\subfigure[Consumer-responsible]{\label{fig:tcc}\includegraphics[width=55mm]{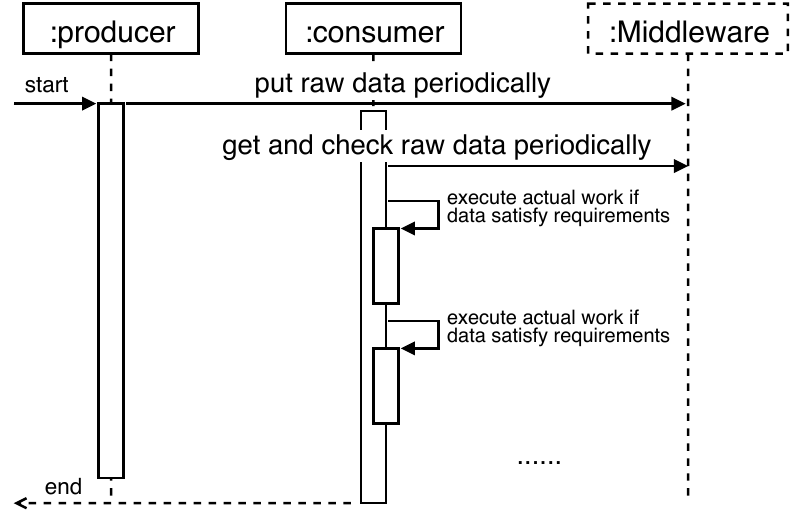}}
\vspace{0.1in}
\subfigure[Middleware-responsible]{\label{fig:mcn}\includegraphics[width=55mm]{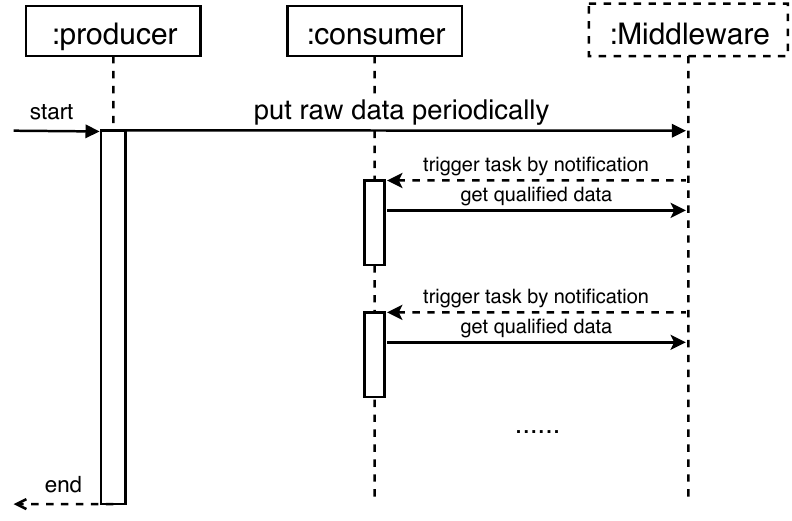}}
\caption{The sequence diagrams of typical dynamic task trigger patterns}
\label{fig:sdtr}
\end{figure*}

This section describes three typical dynamic task trigger patterns. The primary classification standard in this paper is based on the location of a data checking service. For example, we use \textit{Producer-responsible} to represent the case that a data checking service runs at the data producer. The sequence diagrams of the various patterns that are described later-on are represented in Figure~\ref{fig:sdtr}. %It is worth pointing out that the middleware contains different services for each pattern, and all services are initialized properly before the task begins. 

\textbf{Producer-responsible}: This pattern represents the case that the data checking service is integrated with the data producer, and Figure~\ref{fg:tcp} illustrates this pattern in detail. The data producer (such as simulation) generates data at every step, and then the data checking service inspects the content of the data at the end of each step. If the data satisfies the user-defined requirements, the qualified data is sent to the data consumer (such as analytics or visualization) for further processing. With this pattern, all the data received by the consumer are qualified, and the actual work at the data consumer can be triggered directly when data transfer is completed. The advantage of this pattern is to save the workflow execution time by reducing the total amount of data transferred between tasks; however, the data checking service must be coded and compiled together with the simulation program, which increases the complexity of modifying the data checking service and limits the flexibility of changing the data checking techniques dynamically. 

\textbf{Consumer-responsible}: For this pattern, the data checking service is executed at the data consumer. In Figure~\ref{fig:tcc}, the raw data generated by the producer are transferred to the consumer by the I/O service at every step. Afterwards, the content of the data is inspected by the data checking service, which is integrated with the data consumer. The actual work is triggered by the consumer when the data satisfy the requirements. The data checking service can be maintained separately from the simulation or data-producer because of the decoupling, but it requires integration with the data-consumer. In this case, the data-checking operation can be overlapped with the data-generation due to the decoupling between data-producer and data-consumer. The downside for this pattern is that a large amount of the data is transferred between tasks. For instance, even if the data is not interesting, the data must be transferred to consumers to identify if the data is interesting or not, which can become a bottleneck for cases where only few instances of the data are interesting.

\textbf{Middleware-responsible}: For this pattern, the data checking service runs using a separate program in the middleware rather than being integrated with the producer or consumer. As shown in Figure~\ref{fig:mcn}, the data producer puts the raw data into the data I/O service (included in middleware) at each step, and the data analytics are triggered when there is the detection of the qualified data. The instructions about how to start or notify the data analytics can be registered into the middleware before the workflow starts, and the different data analytics can be started according to various data checking results. Compared with other patterns, the instances of data analytics can be started or notified when necessary, such as when there is a detection of the qualified data. 

\section{Evaluation}
\label{sec:experiment}

In this section, we explore how the workflow settings influence the performance of the workflow constructed by different dynamic task trigger patterns. This section mainly compares workflow performance from the perspective of total workflow execution time. Additionally, we also provide insights into which task trigger design strategy works best under user-defined constraints.  %After discussing the evaluation results in detail, we summarize how to choose the design strategies of the dynamic task trigger.

\subsection{Experimental Setup}
\label{sec:setting}

%This section describes the software stack and initial settings of all components used in the evaluation. These components are necessary pieces to construct the dynamic task triggers described in Section~\ref{sec:ttp}.

\textbf{Experiment Applications:} The Gray-Scott simulation and associated analytics discussed at Section~\ref{sec:moti} serve as the data producer and consumer in this evaluation. The BP4 engine of ADIOS2~\cite{lofstead2008flexible} is adopted as the I/O service to transfer data between different tasks. Besides, The middleware~\cite{metadata} supporting the topic matching and the task trigger is adopted as the communication service in this evaluation. This middleware can trigger customized commands when there is match between the published and subscribed topics~\cite{eugster2003many}.

%This evaluation uses the ADIOS2 as the implementation of the I/O service. ADIOS2 offers different types of engines and the various implementations of the engines are hidden under the same $get$/$put$ interfaces. 

%The service of topic matching supports the dynamic task trigger based on the topic-based publish/subscribe interfaces~\cite{eugster2003many}. Specifically, the message called “the topic string” is subscribed to the~\textit{topic matching service} before workflow begins. The message also contains the commands that declare how to trigger the tasks. The topic matching service used in this evaluation are publicly available at.

\textbf{Hardware and Configuration:} We use Amarel supercomputer~\cite{machine} of Rutgers University in this evaluation. The cluster partition used in the evaluation contains 120 nodes and each node is equipped with $28$ Xeon e5-2680v4 (Broadwell) cores and $128$ GB RAM. All tasks are started by the \textit{srun} command and submitted by \textit{Slurm} jobs. There are $8$ MPI processes running in parallel for both the simulation and the analytics. The total number of nodes used in this evaluation is $12$, and the default configurations of the ADIOS2 BP4 engine~\cite{bp4engine} are adopted. \textbf{All the scripts and source code used in evaluation are publicly available at~\cite{dtexp}}. 

%For every process, we set the memory used by per CPU as $1$ GB and the maximum number of the task running on each node as $24$.

\subsection{Metrics and Factors}
\label{sec:variable}

The primary metric evaluated in experiments is the workflow execution time. Specifically, it is the period from the moment that the simulations begin to the moment that all the analytics finish. The workflow execution time is composed of two parts. The first part is the time spent on task execution, such as data producing of simulation and data consuming of analytics. The second part is the time spent on the overhead of the dynamic task trigger and the data transfer between tasks. Since tasks such as simulation in scientific workflow always runs multiple iterations, the execution of different tasks might overlap with each other, and the bottleneck of the workflow also depends on the specific task settings. In this evaluation, we distinguish the task settings according to if the stage of the data generation is the bottleneck of workflow execution (Section~\ref{tasesetting}). For the overhead of the dynamic task trigger and the data transfer, the primary influential factor is the amount of data transferred between tasks. This factor is further affected by the data size generated at each step and the percentage of the qualified data (Section~\ref{exp_ds}). Besides, we also evaluated other factors related to the overhead of the task trigger, such as the number of the triggered analytics (Section~\ref{exp_varana}) and the distribution of the interesting data (Section~\ref{exp_distr}).

\subsection{Results} 
\label{sec:eresults}

\begin{figure}[t]
\centering
\subfigure[]{%\label{fig:b}
\includegraphics[width=60mm]{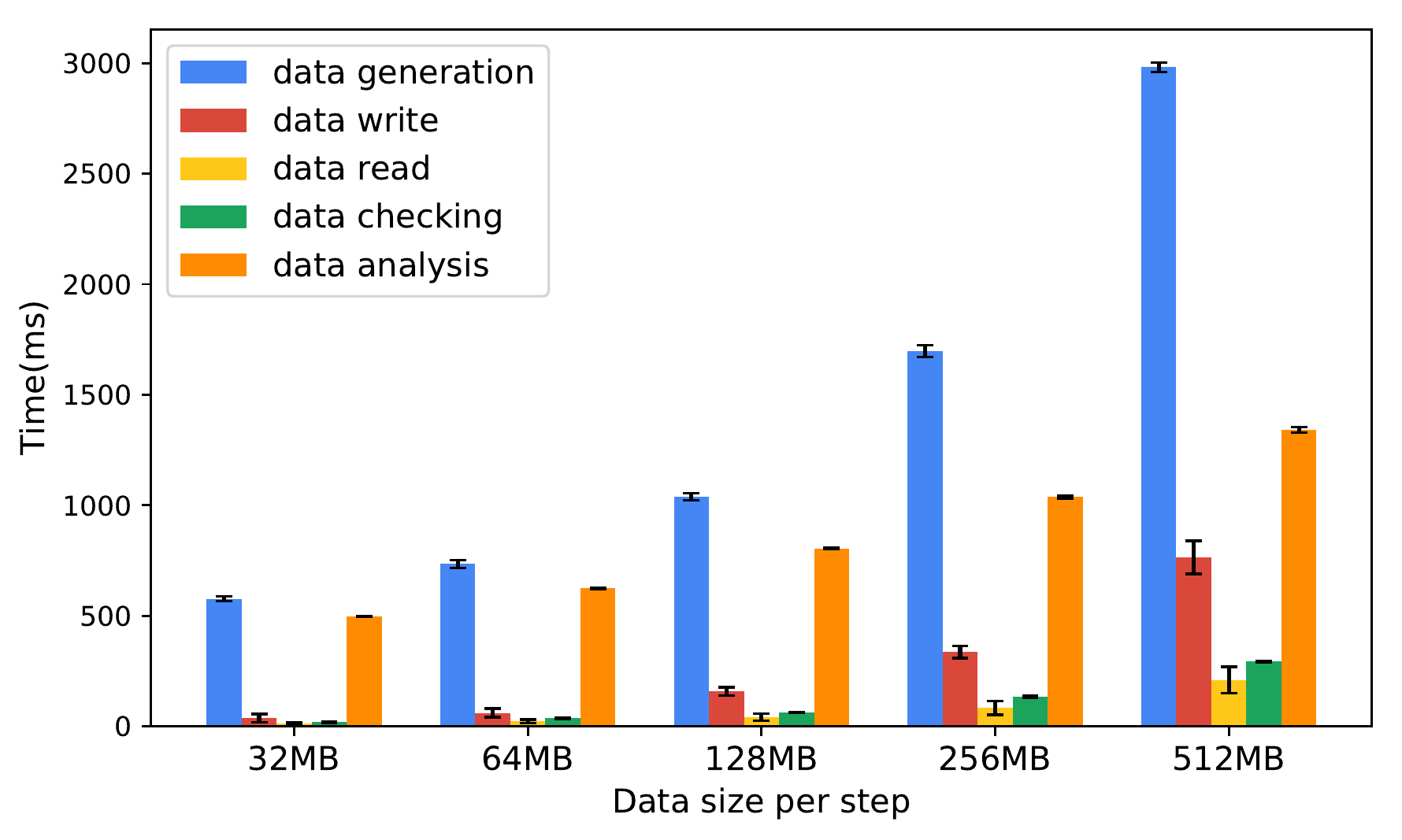}}
\subfigure[] {%\label{fig:a}
\includegraphics[width=60mm]{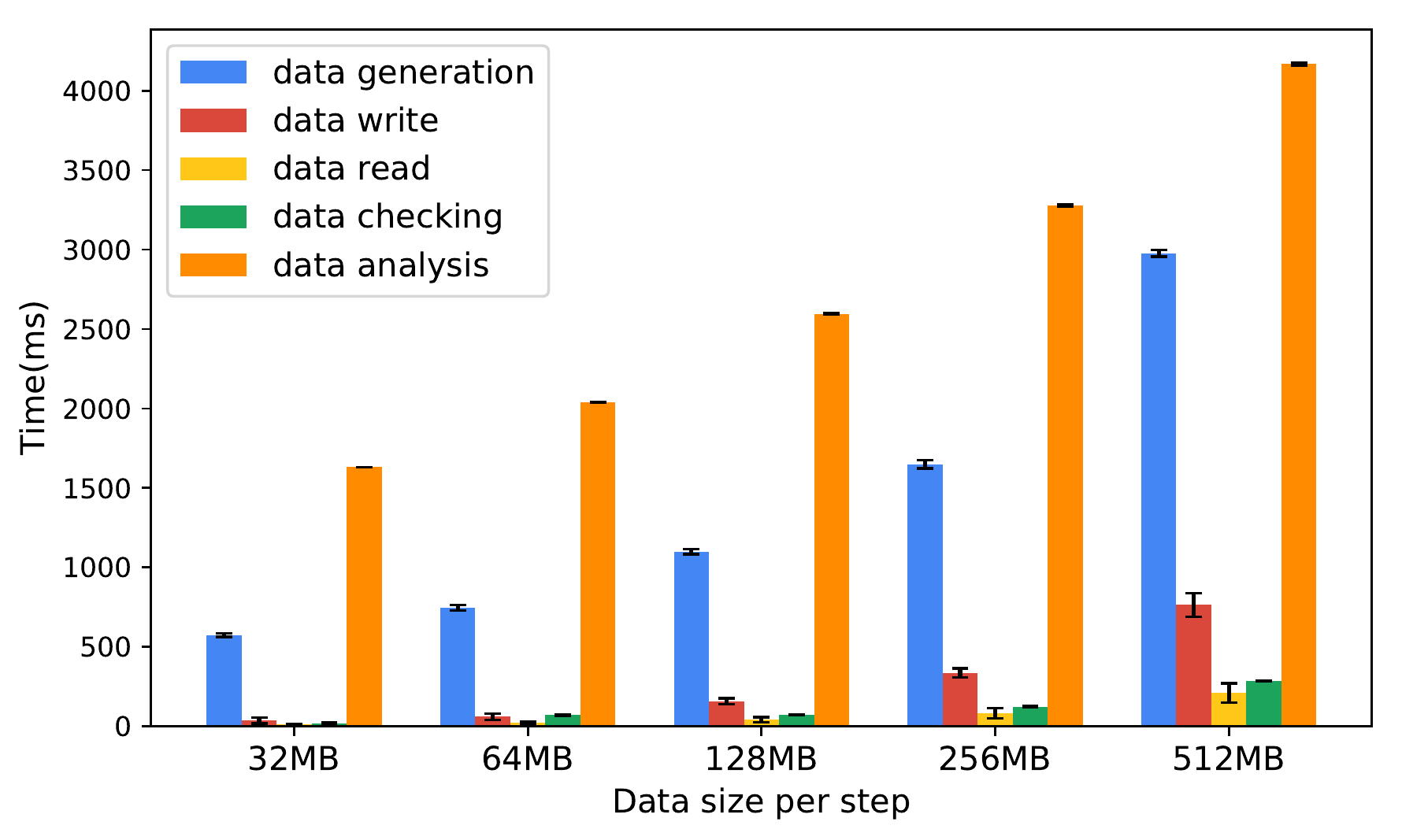}}
\caption{The average time spent on the data generation, data checking, data I/O, and data analysis at each step. The sub-figure (a) shows the case in which the time spent on the data generation is the bottleneck of the workflow. The sub-figure (b) shows the case in which the time spent on the data generation is not the bottleneck of the workflow.}
\label{fg:componentTime}
\centering
\end{figure}

\subsubsection{Typical task settings}\label{tasesetting} The workflow execution time includes the time spent on the task execution and the time spent on the overhead of the data transfer or dynamic task trigger. Before varying specific factors that influence the overhead, it is necessary to illustrate the execution time of tasks used in this evaluation. The process of data generation such as simulation in scientific workflows is usually the compute-intensive task that takes more time than data analytics; however, it is also possible that the data consumer such as analytics takes more time than data generation in particular workflows~\cite{juve2013characterizing}. Both cases need to be considered in this evaluation. As illustrated in Figure~\ref{fg:componentTime}, we calculate the average execution time of different stages included in the workflow. The \textit{x}-axis represents the size of the data generated at each step, and the \textit{y}-axis shows the average execution time (millisecond). In particular, the Figure~\ref{fg:componentTime}(a) illustrates the case in which the time spent on data generation is the bottleneck of the workflow.  Figure~\ref{fg:componentTime}(b) represents the opposite case in which the data generation is not the bottleneck of the workflow. The bottleneck of the workflow in Figure~\ref{fg:componentTime}(b) can be data analytics or data I/O, which depends on specific use cases~\cite{juve2013characterizing}. It is also worth noting that the data checking service is a lightweight analytics compared with other tasks, and more complex data analytics are triggered when the data satisfy the user defined requirements.

\subsubsection{Experiments with various data sizes}\label{exp_ds}

This experiment aims to evaluate how the variation of data size influences the overhead of the workflow execution. There are two factors affecting the size of the data transferred during the workflow execution. The first factor is the size of data generated at each step, and the second factor is the percentage of the qualified data that triggers the data analytics. In this experiment, we choose different combinations of these two factors and compare the execution time of workflow constructed by various dynamic task trigger patterns.

\begin{figure}[t]
\centering
\subfigure[]{%\label{fig:b}
\includegraphics[width=43mm]{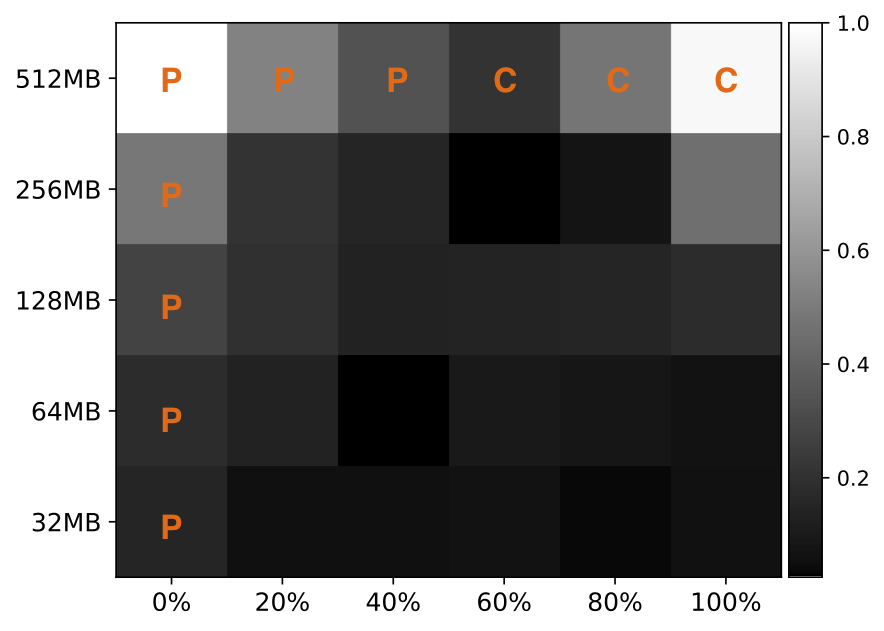}}
\subfigure[] {%\label{fig:a}
\includegraphics[width=43mm]{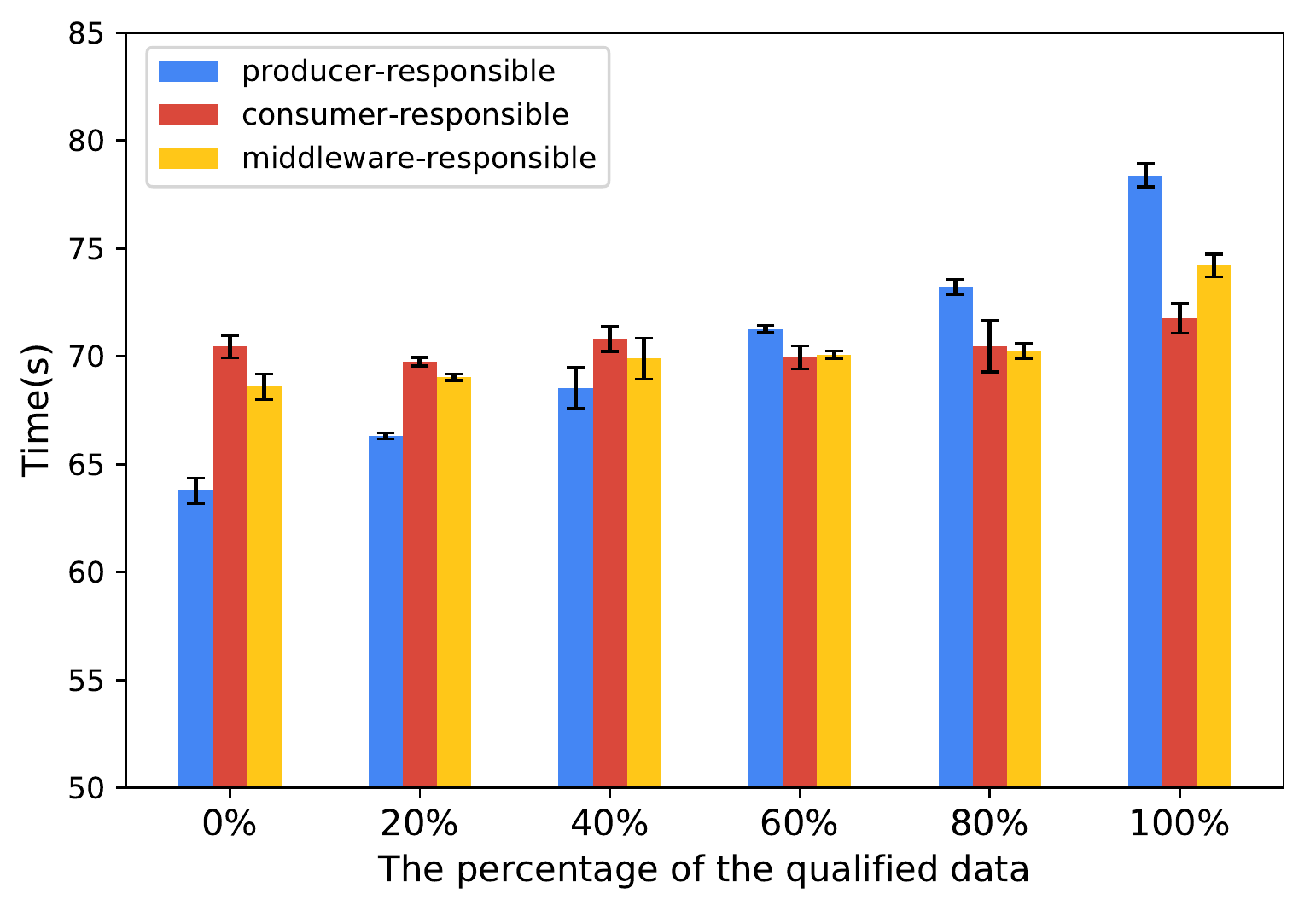}}
\caption{
The colored grid in sub-figure (a) shows how various sizes of data influence the workflow execution time with different dynamic task trigger patterns, and the task settings described in Figure~\ref{fg:componentTime}(a) is adopted in this case. The $x$-axis represents the percentage of the qualified data, and the $y$-axis represents the size of data generated at each step. The gray value at every grid cell represents the degree of distinction of the workflow execution time for different dynamic task trigger patterns. The text shows the pattern with a minimal workflow execution time. Specifically, P represents the Producer-responsible, C represents the Consumer-responsible. The sub-figure (b) shows the detailed results when the data size is $512$MB for each step.}
\label{fg:datasize_sim_greater_ana}
\centering
\end{figure}

\textbf{Case 1:} In this case, we adopt the task settings in which the data generation is the bottleneck of the workflow execution (discussed in Figure~\ref{fg:componentTime}(a)). As results shown in Figure~\ref{fg:datasize_sim_greater_ana}(a), each cell represents the specific combination of the data size and the percentage of the qualified data. The gray value in every cell represents the degree of distinction between the workflow execution time of different patterns. In particular, we use the same configuration to run the workflow constructed by different patterns (discussed at Section~\ref{sec:ttp}). After calculating the average execution time of different patterns, we calculate the standard deviation of these values. This standard deviation value was mapped into the range from $0$ to $1$ and visualized by the gray level in each grid cell. The cell with a lighter color means there is an obvious distinction between different patterns. On the contrary, the cell with the darker color represents there is similar workflow execution time between different patterns. We also labeled the pattern with the minimal execution time on the cell where there is an obvious distinction between different patterns.

One important observation is that the distinction of dynamic task trigger patterns becomes significant only in particular configurations. As illustrated in Figure~\ref{fg:datasize_sim_greater_ana}(a), when the data size is a large number such as $512$MB, the difference between workflow execution time is obvious; however, this difference becomes trivial when the data size is less than $256$MB. This is because the large data size influences the time spent on data I/O, and the overhead of different dynamic task trigger patterns relates to the data I/O.

For a better comparison, we list the the detailed results when the data size at every step is $512$MB. As illustrated in Figure~\ref{fg:datasize_sim_greater_ana}(b), when the percentage of the qualified data is less than $20\%$, the pattern of P (Producer-responsible) is always the optimal solution because most of the data was filtered out and not transferred to other components. It is also worth noting the transition of the optimal pattern with the increase of the qualified data. With the increase of the qualified data, the pattern of C (Consumer-responsible) becomes an optimal solution. This is because checking the data at the simulation (pattern of P) also increases the execution time of the data generation; however, there is an overlap between the data checking and the data generation for the C pattern. When there is a high percentage of the qualified data, it is inefficient to check the data at the producer because most of the data is qualified, and it is more efficient to transfers data to the data consumer directly.

%For the P pattern, the data checking operation is executed at the data producer when the data generation finishes. Since the time spent on data generation is the bottleneck of the workflow in this experiment, the overhead of the data checking and the data I/O is accumulated into the total workflow execution time anyway. However, for the C pattern, only the overhead of data I/O is accumulated into the execution time of the workflow. When most of the data needs to be processed by the consumer (the high percentage of the qualified data), it is more efficient to transfer the data to analytics directly. Since the stage of the data checking and consuming can run concurrently with the stage of the data generation of the next iteration. The process of the data checking is not accumulated into the total execution time because it is overlapped with the execution of the data generation. The execution time for the pattern of M (Middleware-responsible) is similar to the pattern of C, because both patterns execute data checking in concurrent with the data generation.

% TODO, update the explanation above when we update the middleware responsible pattern

\begin{figure}[t]
\centering
\subfigure[]{%\label{fig:b}
\includegraphics[width=43.5mm]{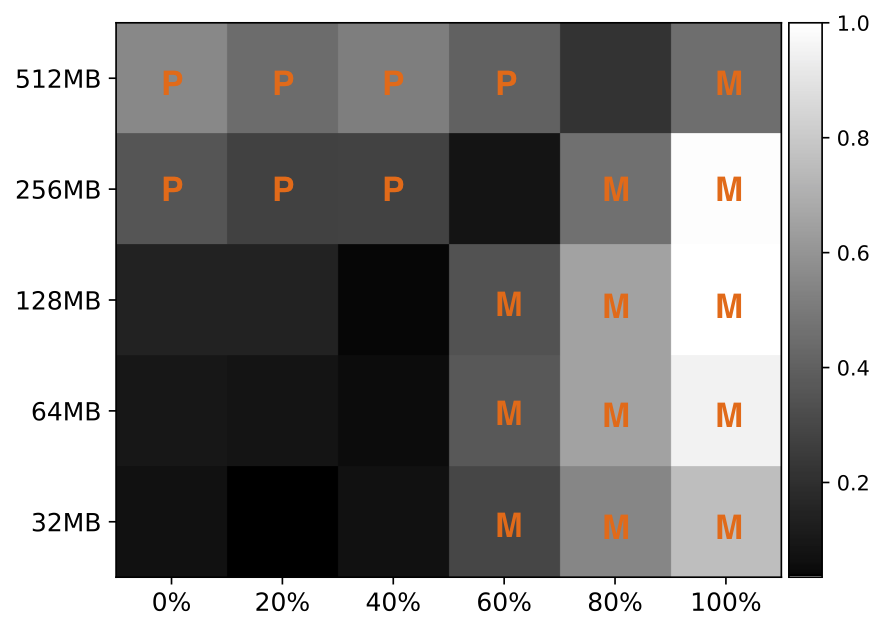}}
\subfigure[] {%\label{fig:a}
\includegraphics[width=43.5mm]{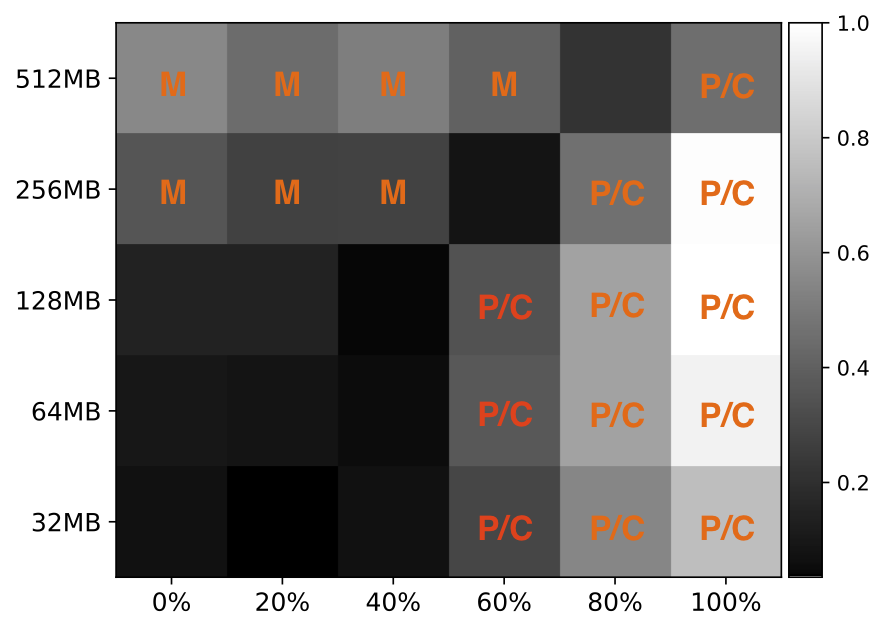}}
\caption{The cell value is calculated in the same way with the Figure~\ref{fg:datasize_sim_greater_ana}(a). The text in sub-figure (a) represents the dynamic task trigger pattern with the minimal workflow execution time, and the text in sub-figure (b) represents the dynamic task trigger pattern with the maximum workflow execution time. In particular, P, C and M represents the pattern of Producer-responsible, Consumer-responsible and Middleware-responsible. P/C represents there is similar workflow execution time between the pattern of P and C.}
\label{fg:datasizesim_less_ana}
\centering
\end{figure}

\textbf{Case 2:} In this case, we adopt the task settings discussed in Figure~\ref{fg:componentTime}(b), in which the bottleneck of the workflow execution is not the time spent on the data generation. The evaluation results are shown in Figure~\ref{fg:datasizesim_less_ana}(a), in particular, the meaning of every cell keeps the same with Figure~\ref{fg:datasize_sim_greater_ana}(a) described at Case 1, and the text in cells represent the patterns with the minimal workflow execution time. In comparison, the patterns with the longest workflow execution time are labeled in Figure~\ref{fg:datasizesim_less_ana}(b). The cells such as ($60\%$, $256$MB) are not labeled by text because the workflow execution times are similar between different patterns with corresponding configurations in these cells. 

One important observation is the distribution of the dark and the light regions. When the percentage of the qualified data is less than $40\%$ and the data size is less than $128$MB, there is no obvious distinction between different dynamic task trigger patterns; however, with the increase of the data size and the percentage of the qualified data, the difference become apparent. It is also worth pointing that there is a significant distinction for different patterns when the percentage of the qualified data is a large number such as $60\%$-$100\%$. Since in this experiment, the time spent on data analytics is longer than data generation (the task settings in Figure~\ref{fg:componentTime}(b)), and there are a large amount of qualified data generated in a relatively short time if the percentage of qualified data is $80\%-100\%$. The pattern of M (Middleware-responsible) can trigger multiple analytics and run them concurrently to fully utilize all available resources to accelerate the workflow execution. This explains why it is the optimal solution when the percentage of qualified data is high.

\begin{figure}
\centering
\subfigure[]{%\label{fig:b}
\includegraphics[width=43.5mm]{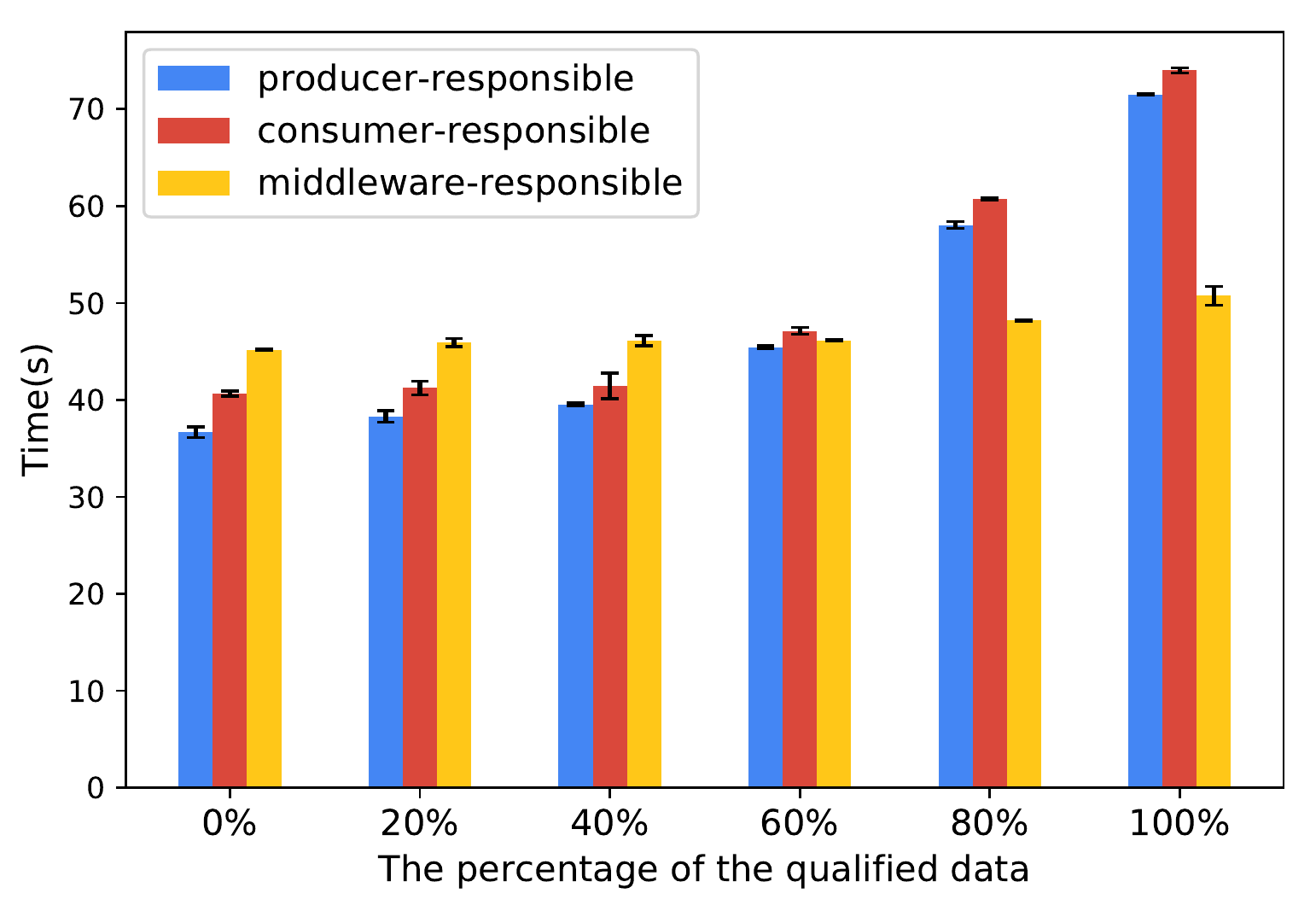}}
\subfigure[] {%\label{fig:a}
\includegraphics[width=43.5mm]{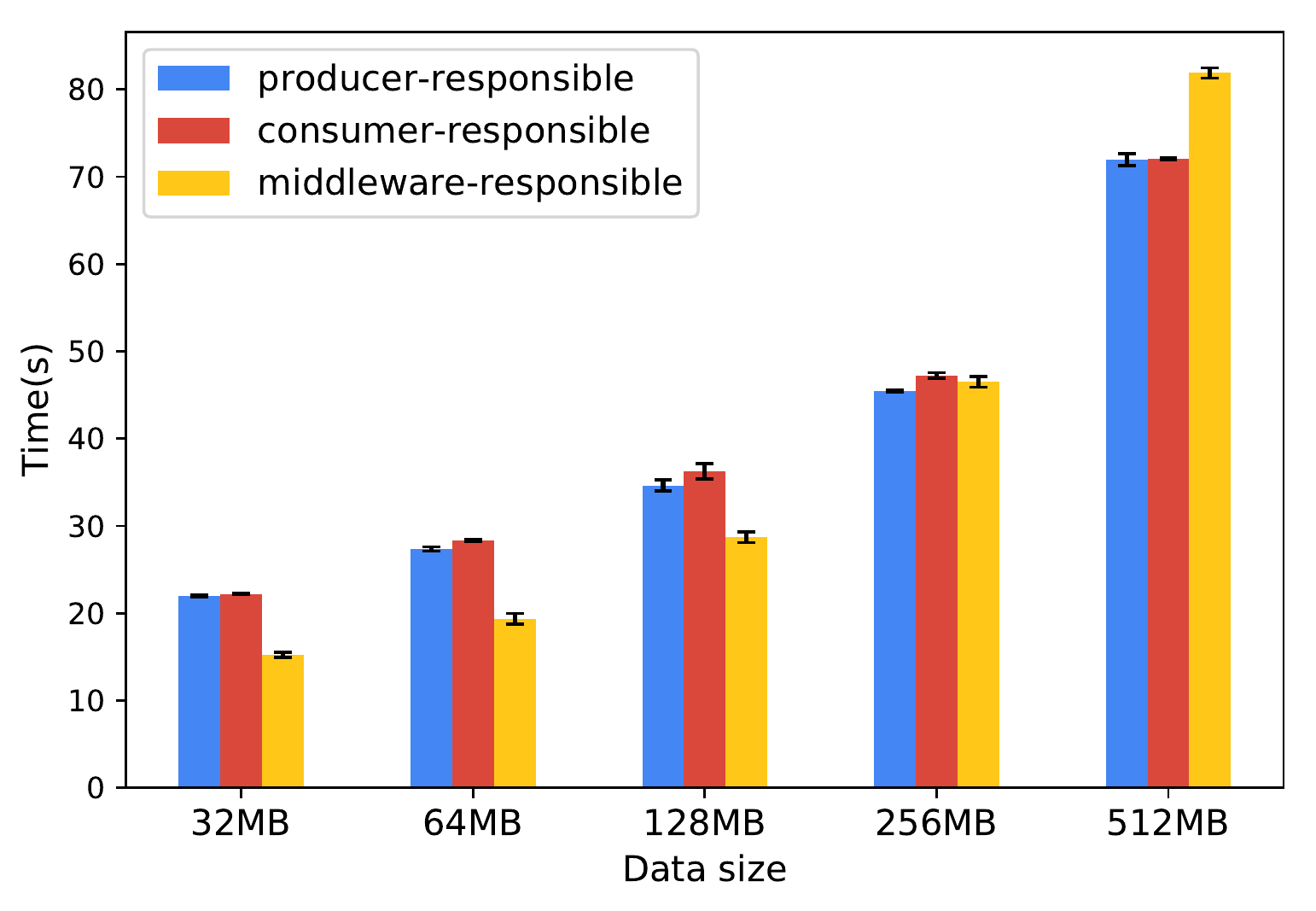}}
\caption{The sub-figure (a) shows the workflow execution time for $256$MB data with various percentages of the qualified data. The sub-figure (b) shows the workflow execution time of $60\%$ qualified data with various sizes of data.}
\label{fg:datasize_cases}
\centering
\vspace{-3mm}
\end{figure}

Another observation is the transition of the optimal solution when the data size is large. For the better comparison, we listed the detailed results when there is $256$MB data generated at each step in Figure~\ref{fg:datasize_cases}(a). Specifically, when the percentage of the qualified data is small, such as $0\%$ or $20\%$, the overhead of the data I/O is the dominant factor that slows down the workflow execution, and the P pattern is preferable than other patterns; however, the benefit of running the analytics concurrently overweight the benefit of decreasing the amount of data transferred between tasks when the percentage of the qualified data is high. This explains the transition of the optimal solution with the increase of the percentage of the qualified data. 
The Figure~\ref{fg:datasize_cases}(b) illustrates the detailed data when the percentage of the qualified data is fixed as the $60\%$, and the data size increases from the $32$MB to $512$MB for each step. When the data size is less than $256$MB, the M pattern is preferable since there is better computing resource utilization; however, with the continuous increase in data size, the overhead of data I/O become significant, and it also increases the workflow execution time for the pattern of M. Although concurrently triggering and executing tasks can fully utilize the computing resources and save the workflow execution time, there is also extra overhead if the task requires more computing resources than available computing resources. This situation is evaluated in the subsequent experiment.

\subsubsection{Experiments with various numbers of analysis tasks}

\label{exp_varana}
\label{taskdepedency}

\begin{figure}[t] 
\centering

\subfigure[]{%\label{fig:b}
\includegraphics[width=42.5mm]{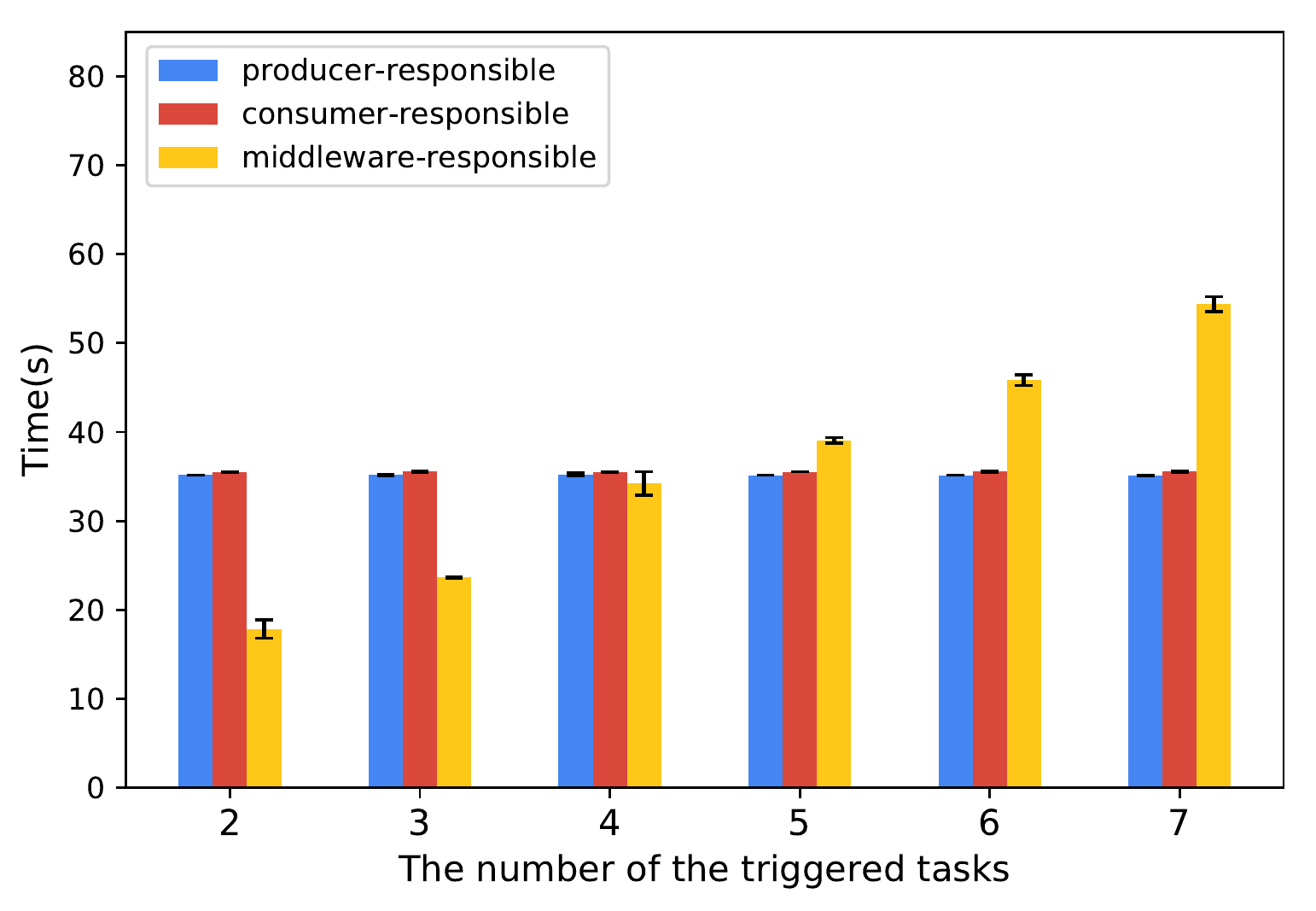}
}
\subfigure[]{%\label{fig:b}
\includegraphics[width=42.5mm]{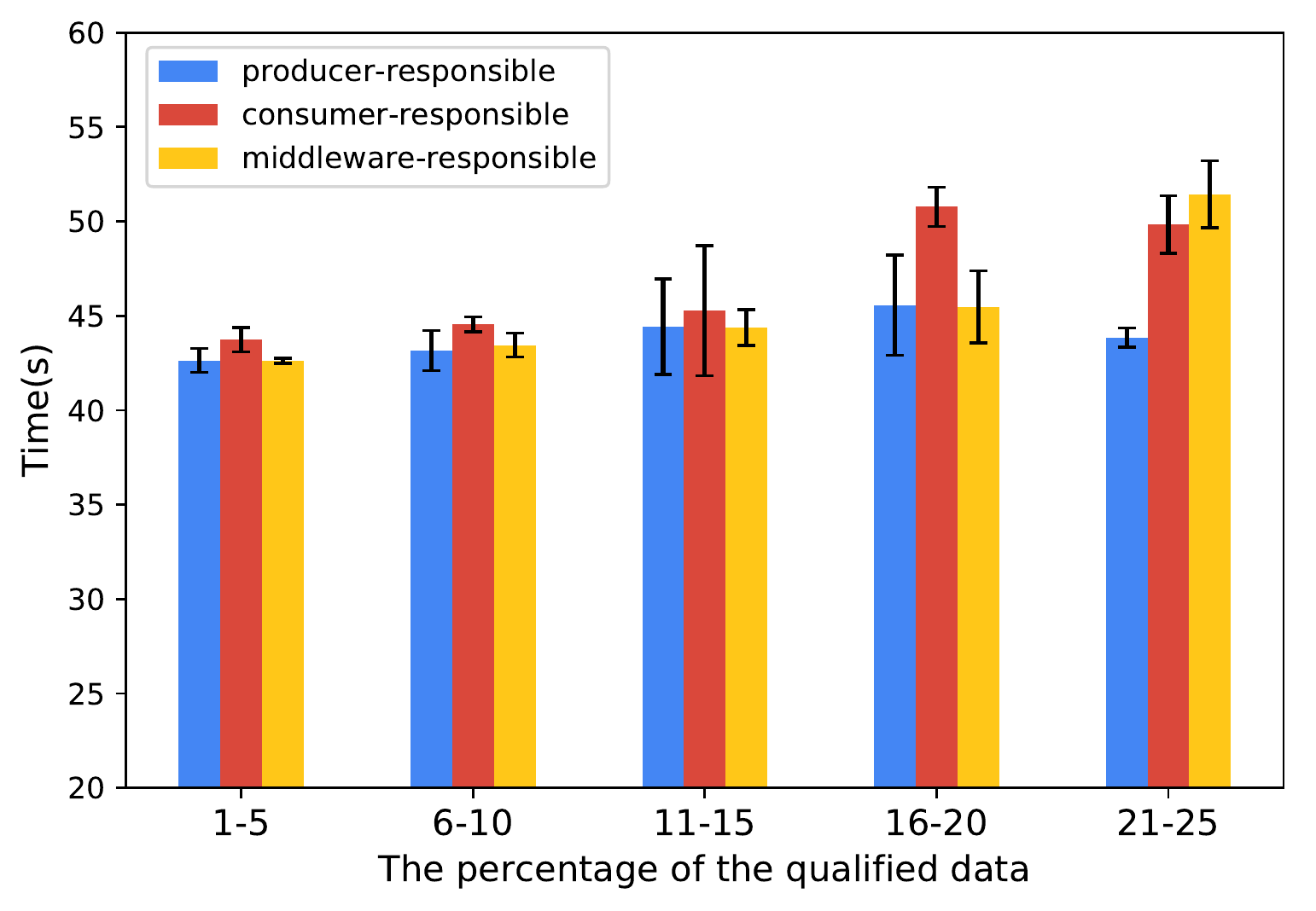}

}
\caption{The sub-figure (a) shows the evaluation with multiple analysis tasks triggered by the data checking service. The sub-figure (b) shows the evaluation with a variant distribution of the qualified data. For example, the $1$-$5$ indicates that the data generated by step $1$-$5$ satisfy the user-defined requirements.}
\label{fg:topoanddist}
\end{figure}

For previous experiments, we assumed there is one type of data analytic associated with the raw data; however, it is also possible that multiple types of data analytics are triggered by data checking services. For example, the different visualizations or analytics might be triggered according to the value range of the data checking results. For this experiment, we aimed to evaluate how the number of analytics influences the workflow execution time. The task setting adopted by this experiment is the case discussed at the Figure~\ref{fg:componentTime}(b). This is because the number of analytics relates to the overhead of the dynamic task trigger, and the workflow execution time is more sensitive to the overhead of dynamic task trigger when the data generation is not the bottleneck of the workflow execution. 

Figure~\ref{fg:topoanddist}(a) shows the workflow execution time with various numbers of analytics. The size of the data generated at each step is $32$MB, and the percentage of the qualified data is $100\%$ in this experiment. It is worth pointing that although in Figure~\ref{fg:datasizesim_less_ana}(a), the optimal solution with the configuration ($100\%$, $32$MB) is the pattern of M, there is a more complicated transition when we varied the number of triggered analytics. In Figure~\ref{fg:topoanddist}(a), when the number of the triggered analytics is small, such as two or three, the optimal solution is the M pattern; however, with an increase in the task number, such as four triggered analytics, the execution time of pattern M increases dramatically and exceeds other patterns. In this experiment, multiple fine granularity analytics are triggered during a short period of time, and these tasks only process the qualified data from a specific step then finish. The method that triggers task and starts to pull qualified data dynamically is advantageous only when the computing resource is available for all the running tasks. Otherwise, the frequent scheduling and starting of analytics can saturate all subscribed resources and increase the extra burden of the workflow execution.

\subsubsection{Experiments with various distributions of qualified data}
\label{exp_distr}
For previous experiments, we assumed the qualified data were distributed evenly among all the steps. For example, we caused qualified data to appear at every $5$ step when there was $20\%$ qualified data within $20$ steps; however, for the use case discussed in~\ref{sec:setting}, the qualified data also tends to appear intensively. For example, there is a high possibility that data from multiple continuous steps satisfy the user-defined requirements. In this experiment, we aimed to evaluate how the distribution of the qualified data influences the workflow execution with different dynamic tasks trigger patterns. Specifically, there were $25$ steps, and $20\%$ of data satisfied the user-defined requirements. The task setting discussed at Figure~\ref{fg:componentTime}(b) is adopted since the workflow execution time is more sensitive to the overhead of the task trigger in this case. 

We varied the indicator value associated with the data checking service to make the qualified data occur between steps 1-5, 6-10, 11-15, etc. The size of the data generated by each step was $128$MB, and the evaluation results are shown in Figure~\ref{fg:topoanddist} (b). It is worth noting that when the qualified data are located in the final section of all the iterations, such as $21$-$25$, the P pattern takes the shortest time. The time saved by decreasing data transfer between tasks plays a major role in reducing total workflow execution time. When the qualified data occur between the first several selections, such as $1$-$5$, there is no apparent distinction between different patterns. This is because all data analysis triggered at first several steps, and they can run concurrently with the stage of data generation from the subsequent iterations.

\subsection{Summary} 
\begin{table*}[t]
\centering
\caption{The summary of the experiment results}
\begin{tabular}{ |l|l|l|l| }
\hline
\multicolumn{2}{ |c| }{Factors} & Task setting A & Task setting B  \\ \hline
\multirow{4}[1]{1.5in}{percentage of the qualified data and the size of the data for each step} & ($\leq\%50$, $\leq128$MB) & P &  similar\\
 & ($\leq\%50$, $>128$MB) & P &P \\
 & ($>\%50$, $\leq128$MB) & similar & M\\
 & ($>\%50$, $>128$MB) & C & M\\ \hline
\multirow{2}{*}{number of analysis tasks} & 2 & P & N \\
 & 7 & P & P\\ \hline
\multirow{2}{1.5in}{distribution of the qualified data} 
 & 5-10 & P & similar \\
 & 20-25 & P & P\\ \hline
\end{tabular}
\label{tb:summary}
\end{table*}

Table~\ref{tb:summary} summarizes the factors that influence the workflow execution time based on the experiment results. The column of \textit{Factors} lists typical configurations evaluated in experiments. The \textit{Task setting A} and \textit{Task setting B} represent initial task settings described by Figure~\ref{fg:componentTime}(a) and Figure~\ref{fg:componentTime}(b), respectively. The preferable dynamic task trigger pattern with the specific configuration is also listed in this table. Three dynamic task trigger patterns (Section~\ref{sec:ttp}) are represented by P, C and M, which means Producer-responsible, Consumer-responsible, and Middleware-responsible, respectively. It is worth noting that values provided in Table~\ref{tb:summary} are based on the experiment setting discussed at Section~\ref{sec:setting}.

% the classification of task settings
According to the evaluation results, the strategies used to decide the dynamic task trigger pattern depend on the type of the workflow tasks. Therefore, it is necessary to distinguish the type of tasks used in workflow before choosing a suitable dynamic task trigger pattern. There are two typical types of task setting evaluated in this work. For the first case (\textit{Task setting A} in Table~\ref{tb:summary}), the time spent on the data generation is the bottleneck of workflow execution. In this case, there is an overlap between the execution of the data generation and data consumption; therefore, the workflow execution time is insensitive to the overhead of the dynamic task trigger. For the second case (\textit{Task setting B} in Table~\ref{tb:summary}), the time spent on the data generation is not the bottleneck of the workflow execution; therefore, the overhead of the dynamic task trigger pattern is influential to the overall workflow execution time. 

The next step is to decide how to compose different components such as data producer, consumer, and middleware into the workflow with dynamic task triggers. One important design consideration is how to execute the data checking operation. The evaluated design options include checking the data at the data producer, data consumer, or separate the middleware. From the perspective of implementation, it is more flexible to implement the data checking service at the data consumer or separate middleware instead of instrumenting the data producer. This is because of the increased decoupling between the data checking operation and the data generation decrease the modification of the simulation program. From the perspective of the performance, it is better to execute data checking service in-line at the data producer if the percentage of the interesting data is relativity low. For example, in Table~\ref{tb:summary}, when the percentage of the qualified data is less than $50\%$, the P pattern is always the optimal solution. This is because checking the raw data at the producer can decrease the transfer of useless data as much as possible. However, if the overhead introduced by data checking overweight the time saved by reducing I/O, such as configuration ($>\%50$, $>128$MB) for the~\textit{Task setting A} in Table~\ref{tb:summary}, it is better to check the data at the consumer or separate middleware.

Another design consideration of the workflow composition with dynamic task triggers is how to start the consumer such as data analytics when there is the detection of the qualified data. One typical design option is to start a data consumer at the beginning of the workflow and poll the qualified data from the data checking service. The actual data analytics is started when it is successful in pulling qualified data. The alternative design option is to trigger data analytics and run it by a separate program when there is the detection of the qualified data. The data analytics are started by dedicated trigger service when necessary. When the time spent on the data generation is the bottleneck (\textit{Task setting A}), or the percentage of the qualified data is relatively low such as less than $50\%$ in Table~\ref{tb:summary}. The workflow execution time is insensitive to the design options of starting the analytics. This is because the time spent on the data trigger is overlapped with the execution of the data generation; however, when the bottleneck is not the data generation (Task setting B in Table~\ref{tb:summary}), it is better to fully utilize the available computing resource to decrease the overhead of the workflow execution. In this case, starting the task dynamically by a separate program is a preferable strategy. By this way, multiple data analytics are triggered and run concurrently in a short period of time to fully utilize the available resources. For example, when the configuration is ($>\%50$, $>128$MB) for the~\textit{Task Setting B} in Table~\ref{tb:summary}, it is better to use pattern M to start the program of the data analytics by separate programs. 

It is also worth pointing that when the triggered task exceeds the capacity of available computing resources, there is an extra overhead for triggering large amount of data analytics. For example, when the number of analytics is $7$ in Table~\ref{tb:summary}, it is better to decrease the task number and increase the workload processed by each task. Furthermore, the distribution of the qualified data also influence the performance of the workflow. As illustrated in Figure~\ref{fg:topoanddist}(b), the workflow execution time is more sensitive to the overhead of dynamic task trigger when the qualified data is distributed among the last several steps of the data generation.

\section{Conclusions and future work}
\label{sec:conclusion}

In this work, we have presented a study during which typical dynamic task trigger patterns of the loosely coupled scientific workflow were compared. The understanding of design considerations of dynamic task triggers and how to choose a suitable pattern in specific workflow settings are essential for workflow design. This work provides three major contributions toward this goal. First, we provided typical dynamic task trigger patterns of loosely coupled dynamic workflows based on the place of the data checking service. Second, we designed and implemented the experiments to evaluate how the workflow settings influence the workflow execution time in different dynamic task trigger patterns. At last, we have summarized the factors that influence the workflow execution time based on the analysis of the experimental results. The discussion of the experiment results provides several insights that guide dynamic workflow design, such as how to execute the data checking service and how to trigger the data consumer.

Although the evaluation in this report reveals the difference between different dynamic trigger patterns, there still lacks a generalized model to illustrates these differences. In the future, we will explore performance models that distinguish different in-situ paradigms. Besides, we will also further explore how to decrease the overhead introduced by the dynamic task trigger. Specifically, we will focus on how to develop efficient middleware to improve the efficiency of the data-driven scientific workflow. One direction is to use the concept of trigger-based in-situ data management~\cite{osti_1493245}. By this design pattern, the configurable in-situ analytics will be integrated into the data management service. Since the in-situ data analytics, such as data checking services, are executed at the place where the data located, the overhead of the in-situ execution will be decreased further. Another work in the future is to explore the dynamic strategies for scientific workflow management. For example, the data polling time can be modified based on current data payload during the progress of the workflow, and the data checking service can be executed at different places based on the changing of the workflow settings.

\section*{Acknowledgements}

The research presented in this paper is supported by the U.S. Department of Energy, Office of Science, Office of Advanced Scientific Computing Research, Scientific Discovery through Advanced Computing (SciDAC) program. This research was conducted as part of the Rutgers Discovery Informatics Institute (RDI$^2$). Thanks to Shouwei Chen and Siyu Liao for proofreading the paper.
%{\color{red}Check with Dr. Parashar!!}

%This research also used resources of the Oak Ridge Leadership Computing Facility, which is a DOE Office of Science User Facility supported under Contract DE-AC05-00OR22725.

\bibliographystyle{unsrt}
\bibliography{main}

\end{document}